\documentclass[a4paper]{article}

\usepackage{RRA4}
\usepackage{epsfig}
\usepackage{calc}
\usepackage{subfigure}
\usepackage{calc}
\usepackage{amssymb}
\usepackage{amstext}
\usepackage{amsmath}
\usepackage{multicol}
\usepackage{caption}
\usepackage{proof}
\usepackage[latin1]{inputenc}

\RRauthor{
Véronique Cortier
\and
Heinrich Hördegen
\and
Bogdan Warinschi}

\RRetitle{Explicit Randomness is not Necessary when Modeling
  Probabilistic Encryption}

\RRtitle{La modélisation du chiffrement probabiliste ne nécessite pas
une représentation explicite de l'aléa}

\newcommand{\verylongrightarrow}[1]             
      {\setlength{\unitlength}{.01in}           
      \begin{picture}(#1,1) \put(0,0){\vector(1,0){#1}} \end{picture}}

\newcommand{\N}{{\mathbb N}}
 \newcommand{\map}[1]{\overline{#1}}

\def\init{\mathsf{init}}    
\def\stop{\mathsf{stop}}    
    
\def\ek{\mathsf{ek}} 
\def\dk{\mathsf{dk}} 
\def\sk{\mathsf{sk}} 
\def\vk{\mathsf{vk}} 
\def\kn{\mathbf{kn}} 

\def\skey{\mathsf{SKey}}    
\def\vkey{\mathsf{VKey}}    
\def\ekey{\mathsf {EKey}}    
\def\dkey{\mathsf {DKey}}    
\def\id{\mathsf {ID}}    
\def\nonce{\mathsf {Nonce}} 
\def\labels{\mathsf {Label}} 
\def\ciphertext{\mathsf {Ciphertext}}    
\def\signature{\mathsf {Signature}}    
\def\term{\mathsf{Term}}    
\def\p{\mathsf{Pair}}    
\newcommand\pair[2]{\langle #1\;  , #2\rangle}    
\newcommand\encrypt[2]{\{#2\}_{#1}}    
\newcommand\sign[2]{[#2]_{#1}}    
    
\def\X{\mathsf X}    
\newcommand\sett[1]{\{#1\}}    
\newcommand{\tf}{T}
\newcommand{\tfs}{T_{\mathit{Sub}}}
\newcommand{\Roles}{\mathsf{Roles}}    
 \newcommand{\Label}{{\cal L}}

\newcommand{\SymbTr}{\mathsf{SymbTr}}  
 
\newcommand{\indexa}[1]{\mathsf{ag}(#1)}
\newcommand{\indexi}[1]{\mathsf{adv}(#1)}

\newcommand{\corrupt}{\mathbf{corrupt}}    
\newcommand{\new}{\mathbf{new}}   
   
\newcommand{\send}{\mathbf{send}}    

\newcommand{\moves}[1]{\xrightarrow{#1}}    
 \newcommand{\sid}{\mathsf{sid}}

\newcommand{\deductl}{\vdash^l}    
\newcommand{\deduct}{\vdash}

\newcommand{\pto}{\to}

\def\c{\mathsf{c}}

\def\SId{\mathsf{SId}}  
\def\sSId{\mathsf{SID}}

\def\Exec{\mathsf{Exec}}

\newcommand{\gstate}[1]{\mbox{$(SId_{#1}, f_{#1}, H_{#1})$}}

\newcommand{\interp}[1]{\mbox{$[\![ #1 ]\!]$}}

\newcommand{\I}{\mathcal{I}}
\newcommand{\LS}{\mbox{$\mathcal{LS}$}}
\newcommand{\db}{\;|\;}

\newcommand{\faLS}[3]{\mbox{$\forall\mathcal{LS}_{#1}(#2).#3\;$}}
\newcommand{\exLS}[3]{\mbox{$\exists\mathcal{LS}_{#1}(#2).#3\;$}}

\newcommand{\Lone}{\mathcal{L}_1}
\newcommand{\Ltwo}{\mathcal{L}_2}

\newcommand{\Sub}{\mathit{Sub}}

\newtheorem{lem}{Lemma}
\newenvironment{lemma}{\begin{lem}}{\end{lem}}

\newtheorem{df}{Definition}
\newenvironment{definition}{\begin{df}}{\end{df}}

\newtheorem{exam}{Example}
\newenvironment{example}{\begin{exam} \begin{rm}}{\end{rm}\end{exam}}

\newenvironment{proof}{\noindent \textit{Proof.}\begin{rm}}{\end{rm}}

\newtheorem{theo}{Theorem}
\newenvironment{theorem}{\begin{theo}\begin{rm}}{\end{rm}\end{theo}}

\RRabstract{
Although good encryption functions are probabilistic, most symbolic models do not capture this aspect explicitly.
A typical solution, recently used to prove the soundness of such
models with respect to computational ones,
is to explicitly represent the dependency of ciphertexts on random coins as labels.

In order to make these label-based models useful, it seems natural to
try to extend the underlying decision procedures and the implementation
of existing tools.
In this paper we put forth a more practical alternative based on the following soundness theorem.
We prove that for a large class of security properties (that includes
rather standard formulations for secrecy and 
authenticity properties), security of protocols in the simpler model
implies security in the label-based model.
Combined with the soundness result of (\textbf{?})
our
theorem enables the translation of security results in unlabeled
symbolic models to computational security.
}

\RRresume{
Bien que de nombreuses fonctions cryptographiques soient probabilistes, la
plupart de modèles symboliques ne prennent pas explicitement en compte cet
aspect.
Pour prouver la correction de
ces modèles par rapport aux modèles computationnels, il est
pourtant souvent nécessaire
de représenter explicitement l'aléa utilisé dans le chiffrement, à
l'aide par exemple d'étiquettes. 

Il semble alors nécessaire d'étendre les procédures de décision sous-jacentes et
l'implémentation des outils existants aux modèles
basés sur des étiquettes.
Dans cet article, nous proposons une alternative plus pratique, basée
sur le théorème de correction suivant. 
Nous prouvons que, pour une grande classe de propriétés de sécurité
(comme les propriétés standards de secret et d'authentification), la
sécurité de protocoles dans un modèle sans étiquettes implique la sécurité dans
les modèles avec étiquettes.
En combinaison avec le résultat de correction de (\textbf{?}), notre
théorème permet de transférer les résultats de sécurité des modèles
symboliques sans étiquettes vers la sécurité computationnelle.
}

\RRkeyword{
Probabilistic encryption, security models, protocol verification, secrecy,
authentication}

\RRmotcle{
Chiffrement probabiliste, modèles de sécurité, vérification des
protocoles, secret, authentification}

\RRprojet{Cassis}

\RRtheme{\THCom \THSym}

\URLorraine

\begin{document}
\makeRR


\section{\uppercase{INTRODUCTION}}

\noindent Designers of mathematical models for computational systems need to
find appropriate trade-offs between two seemingly contradictory
requirements.
Automatic verification (and thus usability) typically requires a high
level of abstraction whereas prediction accuracy requires a high level
of details.
>From this perspective, the use of symbolic models for security
analysis is particularly delicate since it seems  that the inherent
high level of abstraction at which such models operate is not able to
capture all aspects that are relevant to security.
This paper is concerned with one particular such aspect, namely the
use of randomization in the construction of
cryptosystems~\cite{goldwasser84probabilistic}.

A central feature of the computational, complexity-based models
is the ability to capture and reason explicitly about the use of
randomness.
Moreover,  randomness is essential to achieve any meaningful notion of
security for encryption.
In contrast, symbolic models rarely represent randomness directly.
For example, a typical representation for the encryption of message
$m$ under the public key of entity $B$ is the term
$\encrypt{\ek(B)}{m}$.
Notice that the symbolic representation does not capture the
dependency on the randomness used to generate this ciphertext.
While this abstraction may be sufficiently accurate in certain
settings~\cite{MW04}, in some other settings it is not sufficient.

Consider the following flow in some toy protocol:
\[\begin{array}{rcl}
 A& \rightarrow B:& \{m\}_{\ek(B)},\{\{m\}_{\ek(B)}\}_{\ek(B)}
\end{array}\]
To implement this flow,  each occurrence of $\encrypt{\ek(B)}{m}$
is mapped to a ciphertext.
Notice however that the pictorial description does not specify if the
two occurrences of $\encrypt{\ek(B)}{m}$ are equal (created with
identical coins) or different (created with different coins).
In rich enough protocol specification languages disambiguating
constructs as above can be easily done.
For instance, in a language that has explicit assignments, the two
different interpretation for the first message of the protocol can be
obtained as
\[
x:=\encrypt{\ek(B)}{m};
\mathsf{send}(x,\encrypt{\ek(B)}{x})\;\;\;\;\mathrm{and}\;\;\;\;
\mathsf{send}(\encrypt{\ek(B)}{m},\encrypt{\ek(B)}{\encrypt{\ek(B)}{m}})\;\;
\]
Here, each distinct occurrence of $\encrypt{\ek(B)}{m}$ is interpreted
with different randomness.
Other approaches adopt a more direct solution and represent
the randomness used for encryption \linebreak
explicitly~\cite{herzog, abadi01formal, lowe04analysing, CortierW-ESOP05}.
If we write $\encrypt{\ek(B)}{m}^l$ for the encryption of $m$
under the public key of $B$ with random coins $l$, the two
different interpretations of the flow are:
\[
\mathsf{send}(\encrypt{\ek(B)}{m}^{l_1},
\encrypt{\ek(B)}{\encrypt{\ek(B)}{m}^{l_1}}^{l_2})\;\;\;\;
\mathrm{and}\;\;\;\;
\mathsf{send}(\encrypt{\ek(B)}{m}^{l_1},
\encrypt{\ek(B)}{\encrypt{\ek(B)}{m}^{l_2}}^{l_3})
\]
A model that employs labels to capture the randomness used in ciphertexts
(and signatures) has recently been used to establish soundness of
symbolic analysis with respect to computational \linebreak
models~\cite{CortierW-ESOP05}.
Their results are based on an emulation lemma: for protocol
executions, every computational trace can be mapped to a valid symbolic
trace.
The mapping is then used to translate security properties that hold in
the symbolic model to computational analogues.
The next step towards making the soundness result relevant to practice
is to carry out the security proofs using some (semi-)automated tools
for the symbolic model.

However, to the best of our knowledge, none of the popular tools
(ProVerif~\cite{blanchet01}, \linebreak
CASPER~\cite{casper}, Athenta~\cite{athena}, AVISPA~\cite{sysdesc-CAV05}),  offers capabilities for automatically 
reasoning in models that use labels. 
There are at least two solutions to this problem.
One possibility is to enhance the symbolic models that underlie
existing tools.
Unfortunately such a modification would probably require significant
effort that involves adapting existing decision procedures, proving
their correctness, and verifying and modifying thousands of lines of code.

In this paper we put forth and clarify an alternative solution, used
implicitly in~\cite{CortierW-ESOP05}.
The idea is to keep existing tools unchanged, use their underlying
(unlabeled) model to prove security properties, and then show that the
results are in fact meaningful for the model with labels.
The main result of this paper is to prove 
that for a large class of security properties the approach that we
propose is indeed feasible.

We are currently implementing an AVISPA module for computationally
sound automatic proofs based on the results of this paper.

\paragraph{Results.}
We consider the protocol specification language and the execution
 model developed \linebreak in~\cite{CortierW-ESOP05}.
The language is for protocols that use random nonces, public key
encryption and digital signatures, and uses labels to model the
 randomness used by these primitives.
To each protocol $\Pi$ with labels, we naturally associate a protocol
$\map{\Pi}$ obtained by erasing all labels, and extend the
transformation to execution traces.
To each trace $tr$ of $\Pi$ we associate a trace $\map{tr}$ obtained
by erasing labels and we extend this mapping to sets of traces.
The first contribution of this paper is a proof that the
transformation is sound. 
More precisely we prove that if $tr$ is a valid trace
of $\Pi$ (obtained by Dolev-Yao operations) then $\map{tr}$ is a valid
trace of $\map{\Pi}$.
Importantly, this result relies on the fact that the specification
language that we consider does not allow equality tests between
ciphertexts.
We believe that a similar result holds for most (if not all)  protocol
specification languages that satisfy the above condition.
The language for specifying protocols (with and without labels) as
well as the relation between their associated execution models are in
Section~\ref{section-protocol}.

In Section~\ref{section-alogic} we give two logics, $\Lone^l$ and
$\Lone$, that we use to express security properties for protocols with
and without labels, respectively.
Informally, the formulas of $\Lone$ are obtained by removing the
labels from formulas of $\Lone^l$.
Both logics are quite expressive. 
For example, it can be used to express standard formulations for
secrecy and authenticity properties. 

Next we focus our attention on translating security properties between
the two models.
First, notice that the mapping between the model with and that without
labels is not faithful since it looses information regarding
inequality of ciphertexts.
To formalize this intuition we give a protocol $\Pi$ and a formula
$\phi$ such that $\map{\Pi}$ satisfies $\map{\phi}$ (the formula that corresponds to $\phi$
in the model without labels), but for which $\Pi$
does not satisfy $\phi$.
Anticipating, our example indicates that the source of problems is
that $\phi$ may contain equality tests between ciphertexts, and such
tests may not be translated faithfully.
The counterexample is in Section~\ref{section-main}.

The main result of the paper is a soundness theorem.
We show that for a large class of security properties it is possible
to carry out the proof in the model without labels and infer security
properties in the model with labels.
More precisely,  we identify $\Ltwo^l$ and $\Ltwo$, fragments of $\Lone^l$ and
$\Lone$ respectively, such that the following theorem holds.

Consider an arbitrary protocol $\Pi$ and formula $\phi$ in $\Ltwo^l$.
Let $\map{\phi}$ be a formula in $\Ltwo$ obtained by erasing the
labels that occur in $\phi$.  Then,  it holds that:
\begin{displaymath}
   \map{\Pi}\models\map{\phi} \implies \Pi\models\phi
\end{displaymath}
The logics $\Ltwo^l$ and $\Ltwo$ are still expressive enough to
contain  the secrecy and  authentication formulas.
The theorem and its proof are in in Section~\ref{section-main}.

\section{\uppercase{Protocol}}
\label{section-protocol}

In this section we provide the syntax of protocols with labels. 
The presentation is adapted \linebreak from~\cite{CortierW-ESOP05}.
The specification language is similar to the one  of \linebreak
Casrul~\cite{RT01}; it allows parties to exchange messages built from      
identities and randomly generated nonces using public key   
encryption and digital signatures. 
Protocols that do not use labels are obtained straightforwardly. 

\subsection{Syntax}
\label{section-syntax}    
Consider an algebraic signature $\Sigma$ with the following sorts.   
A sort $\id$ for agent identities, sorts $\skey$, $\vkey$, $\ekey$, $\dkey$   
containing keys for signing, verifying, encryption, and decryption  
respectively.        
The algebraic signature also contains sorts $\nonce$, $\labels$, $\ciphertext$,   
$\signature$ and $\p$  for nonces, labels, ciphertexts, signatures
and pair, respectively.    
The sort $\labels$ is used in encryption and signatures to distinguish 
between different encryption/signature of the same plaintext.
The sort $\term$ is a supersort containing all other sorts, except
$\skey$ and $\dkey$.   
There are nine operations: the four operations $\ek,\dk,\sk,\vk$ are defined on   
the sort $\id$ and return the encryption key, decryption key, signing   
key, and verification key associated to the input identity. 
The two operations $\mathsf{ag}$ and $\mathsf{adv}$ are defined on
natural numbers and return labels.
As explained in the introduction, the labels are used to
differentiate between different encryptions (and signatures) of the
same plaintext, created by the honest agents or the adversary.  
 We distinguish between labels for agents and for the adversary since
 they do not use the same randomness. 
The other operations that we consider are pairing, public key    
encryption, and signing.
   



   
We also consider sets of sorted variables $\X=\X.n\cup \X.a\cup
\X.c\cup \X.s$ and $\X^l=\X\cup \X.l$. 
Here, $\X.n, \X.a,\X.c, \X.s, \X.l$ are sets of variables of sort
nonce, agent, ciphertext, signature and labels, respectively.       
The sets of variables $\X.a$ and $\X.n$ are as follows.    
If $k\in\N$ is some fixed constant representing the number of protocol    
participants, w.l.o.g. we fix the set of agent variables to be     
$\X.a=\sett{A_1,A_2,\ldots,A_k}$, and partition the set of nonce   
variables, by the party that generates them.   
Formally:    
$\X.n=\cup_{A\in \X.a}\X_n(A)^{}$ and $\X_n(A) = \{X^j_A   
\mid j\in\N\}.$ 
This partition avoids to specify later, for each role, which variables stand 
for generated nonces and which variables stand for expected nonces.
\medskip 
 
Labeled messages that are sent by participants are specified using   
terms in $\tf^l$ 
\[\begin{array}{lll}
\Label &::= & X.l \mid \indexa{i} \mid \indexi{j}\\ 
\tf^l &::=& \X\mid a \mid \ek(a) \mid \dk(a) \mid \sk(a) \mid \vk(a)
 \mid n(a,j,s) \mid
 \pair{\tf^l}{\tf^l} 
 \mid \encrypt{\ek(a)}{\tf^l}^{\Label} 
 \mid \sign{\sk(a)}{\tf^l}^{\Label}
\end{array}    
\]
where $i,j\in\N $, $a\in \id$, $j,s\in\N$, $a\in \id$.


Unlabeled messages are specified similarly as terms in the algebra
$\tf$ defined by
\[\begin{array}{lll}    
\tf &::=& \X\mid a \mid \ek(a) \mid \dk(a) \mid \sk(a) \mid \vk(a)
  \mid n(a,j,s)  \mid  
 \pair{\tf}{\tf} 
 \mid \encrypt{\ek(a)}{\tf}
 \mid \sign{\sk(a)}{\tf}
\end{array}    
\]
where $a\in \id$, $j,s\in\N$, $a\in \id$. 

A mapping $\map{\cdot}:\tf^l\rightarrow \tf$ from labeled to unlabeled
terms is defined by removing the labels: $\map{\encrypt{m}{k}^{l}} = 
\encrypt{\map{m}}{\map{k}}$, $\map{\sign{m}{k}^{l}} =
\sign{\map{m}}{\map{k}}$, $\map{f(t_1,\ldots,t_n)} =
f(\map{t_1},\ldots,\map{t_n})$ otherwise. The mapping function is
extended to sets of terms as expected.

The individual behavior of each protocol participant is defined by    
a \textit{role} that describes a sequence of message   
receptions/transmissions.
A $k$-party protocol is given by $k$ such roles.    
   
\begin{definition}[Labeled roles and protocols]   
The set $\Roles^l$ of roles for labeled protocol participants is defined by
$\Roles^l=((\sett{\init}\cup\tf^l)\times(\tf^l\cup\sett{\stop}))^*$. 
A $k$-party labeled protocol is a mapping $\Pi:[k]\to\Roles^l$, where 
$[k]$ denotes the set $\{1,2,\ldots,k\}$.
\end{definition}   
Unlabeled roles and protocols are defined very similarly. The mapping
function is extended from labeled protocols to unlabeled protocols as expected.

We assume that a protocol specification is such that    
$\Pi(j)=((l_1^j,r^j_1),(l_2^j,r^j_2),\ldots )$, the $j$'th role in the   
definition of the protocol being executed by player $A_j$.    
Each sequence $((l_1,r_1),(l_2,r_2),\ldots )\in\Roles^l$   
specifies the messages to be sent/received by the party executing the    
role: at step $i$, the party expects to receive a message conforming 
to $l_i$ and returns message $r_i$.    
We wish to emphasize that terms $l_i^j,r_i^j$ are not actual   
messages, but specify how the message that is received and the message that is  
output should look like.
 

\begin{example}    
\label{ex_syntax}   
The Needham-Schroeder-Lowe   
protocol~\cite{lowe96breaking} is specified as follows: there are two   
roles $\Pi(1)$ and $\Pi(2)$ corresponding 
to the sender's and receiver's role.     
\begin{eqnarray*}  
A\rightarrow B: &  & \encrypt{\ek(B)}{N_a,A}\\    
B\rightarrow A: &  & \encrypt{\ek(A)}{N_a,N_b,B}\\    
A\rightarrow B: &  & \encrypt{\ek(B)}{N_b}    
\end{eqnarray*}
\begin{eqnarray*}
\Pi(1) & = & (\init,\encrypt{\ek(A_2)}{X^1_{A_1},A_1}^{\indexa{1}}),\;\;\;\;
    (\encrypt{\ek(A_1)}{X^1_{A_1},X^1_{A_2},A_2}^L,
     \encrypt{\ek(A_2)}{X^1_{A_2}}^{\indexa{1}})\\ 
\Pi(2)& = & (\encrypt{\ek(A_2)}{X^1_{A_1},A_1}^{L_1},
\encrypt{\ek(A_1)}{X^1_{A_1},X^1_{A_2},A_2}^{\indexa{1}}),\;\;\;\;
 (\encrypt{\ek(A_2)}{X^1_{A_2}}^{L_2},\stop)
\end{eqnarray*}
\end{example}   
  
Clearly, not all protocols written using the syntax above are   
meaningful. In particular, some protocols might be not executable.
This is actually not relevant for our result (our theorem also holds
for non executable protocols).

\subsection{Execution Model}    
\label{section-formal}    

We define the execution model only for labeled protocols. The
definition of the
execution model for unlabeled protocols is then straightforward.
 
If $A$ is a variable or constant of sort agent, we define its   
knowledge by   
 $\kn(A) = \{\dk(A),\sk(A)\}\cup \X_n(A)$,  
\textit{i.e.} an agent knows its secret decryption and signing key as
well as    the nonces it generates during the execution.  
The formal execution model is a state transition system. A \emph{global     
state} of the system is given by $(\SId,f,H)$ where $H$ is a set of     
terms of $\tf^l$ representing the messages    
sent on the network and $f$ maintains the     
local states of all session ids $\SId$.     
We represent session ids as tuples of the form    
$(n,j,(a_1,a_2,\ldots,a_k))\in(\N\times\N \times {\id}^k)$,  
where $n\in\N$ identifies 
the session, $a_1,a_2,\ldots,a_k$ are the identities of the parties    
that are involved in the session and $j$ is the index of the role     
that is executed in this session.    
Mathematically, $f$ is a  function   
$f:\SId\to ([\X\pto \tf^l]\times \N\times \N),$
where $f(\mathsf{sid})=(\sigma,i,p)$ is the local state of session $\mathsf{sid}$.    
The function $\sigma$ is a partial instantiation of the variables
occurring in role $\Pi(i)$ and $p\in \N$ is the control point of the
program.  
Three transitions are allowed.    
\begin{itemize}    
\item    
$(\SId,f,H)\moves{\corrupt(a_1,\ldots,a_l)} (\SId,f,\cup_{1\leq j\leq l}\kn(a_j)\cup H)$.
The adversary corrupts parties by outputting a set of identities.   
He receives in return the secret keys corresponding to the identities.    
It happens only once at the beginning of the execution.   
\item The adversary can initiate new sessions:   
 $(\SId,f,H)\moves{\new(i,a_1,\ldots,a_k)} (\SId',f',H')$ where $H'$, $f'$ and $\SId'$    
are defined as follows. Let    
$s = |\SId| +1$,    
 be the session identifier of     
the new session,    
where $|\SId|$ denotes the cardinality of $\SId$. $H'$ is defined by   
 $H'= H$ and $\SId'= \SId\cup\{(s,i,(a_1,\ldots,a_k))\}$.   
The function $f'$ is defined as follows.   
\begin{itemize}    
\item $f'(\sid) = f(\sid)$ for every $\sid\in\SId$.     
\item $f'(s,i,(a_1,\ldots,a_k)) = (\sigma, i, 1)$ where $\sigma$ is a partial function    
$\sigma:\X\pto \tf^l$ and:    
\[\left\{\begin{array}{lll}    
\sigma(A_j) & = a_j & \quad 1\leq j\leq k \\    
\sigma(X^j_{A_i}) & = n(a_i,j,s) & \quad j\in\N    
\end{array}\right.\]  
  \end{itemize} 
We recall that the principal executing the role $\Pi(i)$ is     
represented by  $A_i$ thus, in that role,  every variable    
of the form     
$X^j_{A_i}$ represents a nonce generated by $A_i$.    
   
\item The adversary can send messages:    
$(\SId,f,H)\moves{\send(\sid,m)} (\SId,f',H')$ where $\sid\in\SId$, $m\in\tf^l$,    
 $H'$, and $f'$      
are defined as follows.     
We define $f'(\sid') = f(\sid')$ for every $\sid'\neq\sid$. 
We denote $\Pi(j)=((l^j_1,r^j_1),\ldots,(l^j_{k_j},r^j_{k_j}))$.  
  $f(\sid) = (\sigma, j, p)$ for some $\sigma, j, p$.
There are two cases.     
\begin{itemize}    
\item Either there exists a least general unifier $\theta$ of     
$m$ and $l^j_p\sigma$. Then $f'(\sid) = (\sigma\cup\theta,j,p+1)$     
and $H'=H\cup\{r^j_p\sigma\theta\}$.    
\item Or we define $f'(\sid) = f(\sid)$ and $H'= H$ (the state remains unchanged).    
\end{itemize}    
\end{itemize} 
If we denote by $\sSId = \N\times\N \times {\id}^k$ the set of all sessions ids, 
the set of \emph{symbolic execution traces} is  
$\SymbTr^l \!=\! (\sSId\!\times\! (\sSId\!\to\! ([\X\!\pto\!
  \tf^l]\!\times \!\N\!\times\! \N))\!\times\! 2^{\tf^l})^*$.
The set of corresponding unlabeled symbolic execution traces is denoted by 
$\SymbTr$.
The mapping function $\map{\cdot}$ is extended as follows:
if $tr = (\SId_0,f_0,H_0),\ldots,(\SId_n,f_n,H_n)$ is a trace of $\SymbTr^l$,
$\map{tr} =
(\map{\SId_0},\map{f_0},\map{H_0}),\ldots,(\map{\SId_n},\map{f_n},\map{H_n})\in\SymbTr$
where $\map{\SId_i}$ simply equal $\SId_i$ and 
$\map{f_i}:\sSId\!\to\! ([\X\!\pto\!
  \tf]\!\times \!\N\!\times\! \N))$ with 
$\map{f_i}(\sid) = (\map{\sigma},i,p)$ if $f_i(\sid) = (\sigma,i,p)$
and $\map{\sigma}(X) = \map{\sigma(X)}$.


\begin{figure}[t]    
\begin{tabular}{c}    
\begin{minipage}{3in}    
\[\begin{array}{p{2in}p{2in}p{2in}}
\begin{prooftree}    
\justifies    
S\deductl m    
\using    
m\in S    
\end{prooftree}    
\quad
&    
\begin{prooftree}    
\justifies    
S\deductl b,\ek(b),\vk(b)    
\using    
b\in \X.a   
\end{prooftree}  &
\mbox{Initial knowledge} \\ \\ \\

\begin{prooftree}    
S\deductl m_1 \quad S\deductl m_2    
\justifies    
S\deductl \pair{m_1}{m_2}    
\end{prooftree}    
\quad    
&    
\begin{prooftree}    
S\deductl\pair{m_1}{m_2}    
\justifies    
S\deductl m_i    
\using    
i\in \{1,2\}     
\end{prooftree}  &
\mbox{Pairing and unpairing} \\ \\ \\

\begin{prooftree}    
S\deductl \ek(b) \quad S\deductl m    
\justifies    
S\deductl \encrypt{\ek(b)}{m}^{\indexi{i}}
\using i\in\N    
\end{prooftree}    
\quad    
&    
\begin{prooftree}    
S\deductl \encrypt{\ek(b)}{m}^{l} \quad S\deductl \dk(b)    
\justifies    
S\deductl m    
\end{prooftree} &
\mbox{Encryption and decryption} \\ \\ \\

\begin{prooftree}    
S\deductl \sk(b) \quad S\deductl m    
\justifies    
S\deductl \sign{\sk(b)}{m}^{\indexi{i}} 
   \using i\in\N   
\end{prooftree}    
& 
\begin{prooftree}    
S\deductl \sign{\sk(b)}{m}^{l}  
\justifies    
S\deductl m    
  \end{prooftree} &
\mbox{Signature}
\end{array}    
\]    
\end{minipage} \\ \\  \\ 
\end{tabular}    
\vspace{-6mm}
\caption{Deduction rules.} 
\vspace{3mm} 
\hrule  
\label{figure-deducta} 
  \vspace{-7mm}
\end{figure}

The adversary intercepts messages between honest participants and
computes new messages using the deduction relation
$\deductl$ defined in Figure~\ref{figure-deducta}. 
Intuitively, $S\deductl m$ means that the adversary is able to compute the
message $m$ from the set of messages $S$.    
All deduction rules are rather standard with the exception of the last
one:
The last rule states that the adversary can recover the corresponding
message out of a given signature.
This rule reflects capabilities that do not contradict the standard
computational security definition of digital signatures, may
potentially be available to computational adversaries and are
important for the soundness result of~\cite{CortierW-ESOP05}.


Next, we sketch the execution model for unlabeled protocols. 
As above, the execution is based on a deduction relation $\deduct$
that captures adversarial capabilities. 
The deduction rules that define $\deduct$ are obtained from those of
$\deduct^l$ (Figure~\ref{figure-deducta}) as follows. 
The sets of rules \textit{Initial knowledge} and \textit{Pairing
  and unpairing} in are kept unchanged (replacing $\deductl$ by $\deduct$, of course). 
For encryption and signatures we suppress the labels  $\indexi{i}$ and
$l$ in the encryption function $\{\_\}^\__\_$ and the signature
function
$\sign{\_}{\_}^\_$ for rules
\textit{Encryption and   decryption} and rules \textit{Signature}.
That is, the rules for encryption are: 
\begin{displaymath}
\begin{prooftree}    
\quad 
S\deduct \ek(b) \quad S\deduct m    
\justifies    
S\deduct \encrypt{\ek(b)}{m}
\end{prooftree}    
\quad 
\quad
\begin{prooftree}    
S\deduct \encrypt{\ek(b)}{m} \quad S\deduct \dk(b)    
\justifies    
S\deduct m    
\end{prooftree}    
\quad    
\end{displaymath}
and those for signatures are: 
\begin{displaymath}
\begin{prooftree}    
S\deduct \sk(b) \quad S\deduct m    
\justifies    
S\deduct \sign{\sk(b)}{m}
\end{prooftree}    
\quad 
\quad
\begin{prooftree}    
S\deduct \sign{\sk(b)}{m}
\justifies    
S\deduct m    
  \end{prooftree}
   \quad 
\end{displaymath}

We use the deduction relations to characterize the set of valid
execution traces. 
We say that the trace $(\SId_1,f_1,H_1), \ldots, (\SId_n,f_n,H_n)$  is
\emph{valid} if the messages sent by the adversary can be computed by
Dolev-Yao operations. 
More precisely, we require that in a valid trace whenever
$(\SId_i,f_i,H_i)\moves{\send(s,m)} (\SId_{i+1},f_{i+1},H_{i+1})$, we
have $H_i\deductl m$.   
Given a protocol $\Pi$, the set of valid symbolic execution traces is
denoted by $\Exec(\Pi)$.  
The set $\Exec(\map{\Pi})$ of execution traces in the model without
labels  is defined similarly. 
We thus require that every sent message $m'$ satisfies 
$\map{H_i} \deduct m'$.  


\begin{example}   
Playing with the Needham-Schroeder-Lowe protocol described in    
Example~\ref{ex_syntax}, an adversary can corrupt an agent $a_3$,
start a new session  for the second role with players $a_1,a_2$ and
send the message     
$\encrypt{\ek(a_2)}{n(a_3,1,1),a_1}^{\indexi{1}}$ to the player of the second role.   
The corresponding valid trace execution is:   
\begin{multline*}   
\!\!\!\!\!
(\emptyset,f_1,\emptyset)\moves{\corrupt(a_3)}\left(\emptyset,f_1,\kn(a_3)\right)   
\moves{\new(2,a_1,a_2)}\\ \left(\{\sid_1\},f_2,\kn(a_3)\right)
\moves{\send(\sid_1,\encrypt{\ek(a_2)}{n_3,a_1}^{\indexi{1}})} \\
\left(\{\sid_1\},f_3,\kn(a_3)\cup
\{\encrypt{\ek(a_1)}{n_3,n_2,a_2}^{\indexa{1}}\}\right),   
\end{multline*}   
where    
$\sid_1 = (1,2,(a_1,a_2))$, $n_2 = n(a_2,1,1)$, $n_3 = n(a_3,1,1)$, and $f_2,f_3$ are defined as follows:   
$f_2(\sid_1) = (\sigma_1,2,1)$, $f_3(\sid_1) = (\sigma_2,2,2)$ where   
$\sigma_1(A_1) = a_1$, $\sigma_1(A_2) = a_2$,   
$\sigma_1(X^1_{A_2}) = n_2$,   
and $\sigma_2$ extends $\sigma_1$ by $\sigma_2(X^1_{A_1}) = n_3$ and
$\sigma_2(L_1) = \indexi{1}$.   
\end{example}   









\subsection{Relating the labeled and unlabeled execution models}
\label{sec:relating}
First notice that by induction on the deduction rules, it can be
easily shown that whenever a message is deducible, then the corresponding
unlabeled message is also deducible. 
Formally, we have the following lemma. 
\begin{lemma}
\label{lem1}
$S \deductl m \Rightarrow \map{S} \deduct \map{m}$
\end{lemma}
Note that our main result holds for any deduction rules provided this
lemma holds. 

Based on the above property  we show that whenever a trace
corresponds to an execution of a protocol, the corresponding unlabeled
trace corresponds also to an execution of the corresponding unlabeled
protocol. 
\begin{lemma}
\label{lem4}
$
tr \in \Exec(\Pi) \Rightarrow \map{tr} \in \Exec(\map{\Pi}).
$
\end{lemma}

\begin{proof}
The key argument is that only pattern matching is
performed in protocols and when a term with labels matches some
pattern, the unlabeled term matches the corresponding unlabeled
pattern. The proof is done by induction on the length of the trace. 

\begin{itemize}

\item Let $tr = \gstate{0}$, where $\SId_0$ and $H_0$ are empty sets.
We have $\map{H_0} = H_0$. $f_0$ is defined nowhere, and so
is $\map{f_0}$. Clearly, $\map{tr} = (\SId_0, \map{f_0}, \map{H_0})$
is in $\Exec(\map{\Pi})$.

\item 
Let $tr \in \Exec(\Pi)$, $tr = e_0, ..., e_n = \gstate{0}, ..., \gstate{n}$, such that
$\map{tr} \in \Exec(\map{\Pi})$. 
We have to show that if $tr' = tr, \gstate{n+1} \in \Exec(\Pi)$, then
we have $\map{tr'} \in \Exec(\map{\Pi})$.
There are three possible operations.
\begin{enumerate}
\item $corrupt(a_1, ..., a_k)$. It means that $tr = \gstate{0}, \gstate{1}$.
In this case, we have $\SId_1 = \SId_0 = \emptyset$, $f_1 = f_0$ and
$H_1 = H_0 \cup \bigcup_{1 \leq i \leq k} \kn(a_i)$.
We can conclude that 
$\map{tr} = (\SId_0, \map{f_0}, \map{H_0}),(\SId_1, \map{f_1}, \map{H_1})$ is
in $\Exec(\map{\Pi})$, because there are no labels in $H_1$ and
$f_1$ is still not defined.

\item $new(i, a_1, ..., a_k)$. No labels are involved in this operation.
The extension made to $f_n$ is the same as is made to $\map{f_n}$.
Neither $H_n$ nor $\map{H_n}$ are modified.
$\map{tr'} = \map{tr}, (\SId_{n+1}, \map{f_{n+1}}, \map{H_{n+1}})$ is a valid trace.

\item $send(s, m)$.

First, we have to be sure that if $m$ can
be deduced from $H_n$, then $\map{m}$ can be deduced from $\map{H_n}$.
This is Lemma~\ref{lem1}.

Note that $\SId_n = \SId_{n+1}$ thus $\map{\SId_n} = \map{\SId_{n+1}}$.
Let $f_n(s)=(\sigma, i, p)$ and 
$\Pi(i) = (..., (l_p, r_p), ...)$. We have two cases.
\begin{itemize} 
\item Either there is a substitution $\theta$ with $m = l_p\sigma\theta$.
Then  $f_{n+1}(s) = (\sigma \cup \theta, i, p+1)$.  
Thus $\map{f_n}(s) = (\map{\sigma}, i, p)$ and 
$\map{f_{n+1}}(s) = (\map{\sigma} \cup \map{\theta}, i, p+1)$.
By induction hypothesis, $\map{tr}$
is a valid trace. From $m = l_p\sigma\theta$ follows
$\map{m} = \map{l_p} \map{\sigma}\map{\theta}$.
We conclude that $\map{tr}, (\SId_{n+1}, \map{f_{n+1}}, \map{H_{n+1}}) =
\map{tr'}$ is a valid trace, thus a member of $\Exec(\map{\Pi})$.

\item Or no substitution $\theta$ with $m = l_p\sigma\theta$ exists.
Then $tr' = e_0, ..., e_n, e_{n+1}$ with $e_n = e_{n+1}$.
We must show that it is always possible to construct a message 
$m' \in \tf$, such that there exists no substitution $\theta'$ with 
$m' = \map{l_p} \map{\sigma} \theta'$. Then, from the validity of $tr'$
and $\map{tr}$ we can deduce the validity of $\map{tr'}$, because
$\map{e_n} = \map{e_{n+1}}$.

Either there exists no substitution $\theta'$ such that $\map{m} = \map{l_p}
\map{\sigma} \theta'$. In that case, we choose $m' = \map{m}$.

Or let $\theta'$ be a substitution such that $\map{m} = \map{l_p}
\map{\sigma} \theta'$. Then the matching for $m$ fails because of
labels. This can be shown by contradiction.
Assume $m$ contain no label, \textit{i. e.} $m$ does not contain subterms
of the form $\{t\}^l_{\ek(a_i)}$ or $[t]^l_{\sk(a_i)}$, $t \in \tf$. In that
case, we have $\map{m} = m$ by definition. 
>From $\map{m} = \map{l_p}
\map{\sigma} \theta'$, we deduce that $m = l_p\sigma\theta'$, contradiction.


We deduce that $\map{m}$ contains some subterm of the form
$\{t\}_{\ek(a_i)}$ or
$[t]_{\sk(a_i)}$. The fact $\map{m} =
\map{l_p} \map{\sigma} \theta'$ implies that $\map{l_p}$ has to
contain one of the following subterms: 
$\{t'\}_{\ek(A_i)}$, $[t']_{\sk(A_i)}$ with $t' \in T$ or,
a variable of sort ciphertext or signature.

Then, we choose $m' = a$ for some agent identity $a\in\X.a$.
The term $a$ is 
deducible from $\map{H_n}$. 
Now, the matching of $m'$ with $\map{l_p}$  always fails, either
because of the
encryption or signature occurring in 
$\map{l_p}$ or because of type mismatch for a variable of type
ciphertext or signature in $\map{l_p}$. 
\end{itemize}
\end{enumerate}
\end{itemize}

\end{proof}

\section{\uppercase{A logic for security properties}}
\label{section-alogic}
In this section we define a logic for specifying security properties. 
We then show that the logic is quite expressive and, in particular,
it can be used to
specify rather standard secrecy and authenticity properties.  

\subsection{Preliminary definitions}
\label{section-Preliminary}

To a trace $tr = e_1, ..., e_n = \gstate{1}, ..., \gstate{n}
\in \SymbTr$ we associate its set of indices $\I(tr) = \{i \db e_i
\mbox{ appears in the trace } tr \}$.

We also define the set of local states $\LS_{i,p}(tr)$ for role $i$ at step
$p$ that appear in
trace $tr$ by
$
\LS_{i, p}(tr) = \{(\sigma, i, p) \db \exists s \in \SId_k, k \in \I(tr),
\mbox{ such that } f_k(s) = (\sigma, i, p) \}.
$
\medskip

We assume an infinite set $\Sub$ of meta-variables for substitutions.
We extend the term algebra to allow substitution application. 
More formally, let $\tfs^l$ be the algebra defined by:
\[\begin{array}{lll}
\Label &::= & \varsigma(x_l) \mid \indexa{i} \mid \indexi{j} \\ 
\tfs^l &::=&  \varsigma(x)  \mid a \mid \ek(a) \mid \dk(a) \mid \sk(a)
\mid \vk(a) \mid
 \pair{\tfs^l}{\tfs^l} 
 \mid \encrypt{\ek(a)}{\tfs^l}^{\Label} 
 \mid \sign{\sk(a)}{\tfs^l}^{\Label} 
\end{array}    
\]
where $x_l\in\X.l$, $\varsigma\in\Sub$,  $i,j\in\N$, $x\in\X$, $a\in
\id$. 
The unlabeled algebra $\tfs$ is defined similarly. 
The mapping function between the two algebras is defined by:
$\map{\varsigma(x)}= 
\varsigma(x)$, $\map{\encrypt{m}{k}^{l}} =
\encrypt{\map{m}}{\map{k}}$, $\map{\sign{m}{k}^{l}} =
\sign{\map{m}}{\map{k}}$, $\map{f(t_1,\ldots,t_n)} =
f(\map{t_1},\ldots,\map{t_n})$ otherwise.

\subsection{Security Logic}
\label{section-security}

In this section we describe a logic for security properties. 
Besides standard propositional connectors, the logic has a predicate to
specify honest agents, equality tests between terms, and existential
and universal quantifiers over the local states of agents. 
\begin{definition}
The formulas of the logic $\Lone^l$ are defined as follows:
\[
\begin{array}{lll}
F(tr) & ::= & NC(tr, t_1) \db (t_1 = t_2) \db \neg F(tr) \db 
F(tr) \wedge F(tr) \db  F(tr) \vee F(tr) \db\\
& & \faLS{i, p}{tr}{\varsigma}F(tr) \db \exLS{i, p}{tr}{\varsigma}F(tr)
\end{array}
\]
where $tr$ is a parameter of the formula, $i,p\in \N$,
$\varsigma\in\Sub$,
$t_1$ and $t_2$ are terms of $\tfs^l$.
Note that formulas are parametrized by a trace $tr$.
As usual, we may use $\phi_1\rightarrow\phi_2$ as a shortcut for $\neg\phi_1\vee\phi_2$.
\end{definition}

We similarly define the corresponding unlabeled logic $\Lone$: the tests $(t_1 =
t_2)$ are between unlabeled terms $t_1,t_2$ over $\tf_\mathit{sub}$.
The mapping function $\map{\cdot}$ is extended as expected. In
particular $\map{NC(tr, t)} = NC(\map{tr}, \map{t})$, $\map{(t_1 =
  t_2)} = (\map{t_1} = \map{t_2})$, 
$\map{\faLS{i, p}{tr}{\varsigma}F(tr)} = \faLS{i, p}{\map{tr}}{\varsigma}\map{F(tr)}$
and $\map{\exLS{i, p}{tr}{\varsigma}F(tr)} = 
\exLS{i, p}{\map{tr}}{\varsigma}\map{F(tr)}$.

Here, the predicate $NC(tr, t)$ of arity 2 is used to specify
non corrupted agents. 
The quantifications $\faLS{i, p}{tr}{\varsigma}$
and $\exLS{i, p}{tr}{\varsigma}$ are over the local states in the
trace that correspond to agent $i$ at control point $p$.
The semantics of our logic is defined for \emph{closed} formula as
shown in Figure~\ref{interpretation}.
\begin{figure*}[t]
\begin{center}
\begin{eqnarray*}
\interp{NC(tr, t)} & = & 
\left\{
    \begin{array}{l l}
  1 &   \mbox{if } t\in\id  \mbox{ and $t$ does not appear
  in a corrupt action, \textit{i.e.} }
\\ & tr=e_1, e_2, ..., e_n \mbox{ and }\\ & \forall a_1,\ldots,a_k
  \mbox{, s.t.  } e_1\moves{\corrupt(a_1,\ldots,a_k)}e_2, t\neq a_i
 \\
  0 &  \mbox{otherwise} 
    \end{array} \right.
\\
\interp{(t_1 = t_2)} & = & \left\{
    \begin{array}{l l}
  1 &   \mbox{if } t_1=t_2  \mbox{ (syntactic equality)}\\
  0 &  \mbox{otherwise} 
    \end{array} \right.
\\
\interp{\neg F(tr)} & = & \neg \interp{F(tr)}\\
\interp{F_1(tr)\wedge F_2(tr)} & = & \interp{F_1(tr)}\wedge \interp{F_2(tr)}\\
\interp{F_1(tr)\vee F_2(tr)} & = & \interp{F_1(tr)}\vee \interp{F_2(tr)}\\
\interp{\faLS{i, p}{tr}{\varsigma}F(tr)} & = & \left\{  
   \begin{array}{l l}
  1 &   \mbox{if } \forall (\theta, i, p) \in \LS_{i, p}(tr), \mbox{ we have }
  \interp{F(tr)[\theta/\varsigma]} = 1,  \\
  0 &  \mbox{otherwise}. 
   \end{array} \right.
\\
\interp{\exLS{i, p}{tr}{\varsigma}F(tr)} & = & \left\{  
   \begin{array}{l l}
  1 &   \mbox{if } \exists (\theta, i, p) \in \LS_{i, p}(tr), \mbox{ s.t. }
  \interp{F(tr)[\theta/\varsigma]} = 1,  \\
  0 &  \mbox{otherwise}. 
   \end{array} 
\right.
\end{eqnarray*}
\vspace{-6mm}
\caption{Interpretation.} 
\vspace{3mm} 
\hrule  
\label{interpretation} 
  \vspace{-7mm}
\end{center}
\end{figure*}

Next we define when a protocol $\Pi$ satisfies a formula
$\phi\in\Lone^l$.  The definition for the unlabeled execution model is
obtained straightforwardly. 
Informally, a protocol $\Pi$ satisfies $\phi$ if $\phi(tr)$ is true
for all traces $tr$ of $\Pi$.  Formally:
\begin{definition}
Let $\phi$ be a formula and $\Pi$ be a protocol. We say that $\Pi$
satisfies security property $\phi$, and write $\Pi \models \phi$ if
for any trace $tr\in\Exec(\Pi)$, $\interp{\phi(tr)}=1$. 
\end{definition}
Abusing notation, we occasionally write $\phi$ for the set $\{tr\mid
\interp{\phi(tr)}=1\}$.  
Then,  $\Pi\models\phi$ precisely when $\Exec(\Pi)\subseteq\phi$.
















\subsection{Examples of security properties}
\label{section-Examples}
In this section we exemplify the use of the logic by specifying
secrecy and authenticity properties.

\subsubsection{A secrecy property}

Let $\Pi(1)$ and $\Pi(2)$ be the sender's and receiver's role of a two-party
protocol. 
To specify our secrecy property we use a standard encoding.
Namely, we add a third role to the protocol, $\Pi(3) = (X^1_{A_3},
stop)$, which can be seen as some sort of witness.

Informally, the definition of the secrecy property $\phi_s$ states that,
for two non corrupted agents $A_1$ and $A_2$, where $A_1$ plays role
$\Pi(1)$ and $A_2$ plays role $\Pi(2)$, a third agent playing role 
$\Pi(3)$ cannot gain any knowledge on nonce $X^1_{A_1}$ sent by
role $\Pi(1)$.
\[
\phi_s(tr) =\faLS{1,1}{tr}{\varsigma} \faLS{3,2}{tr}{\varsigma'}
[ NC(tr, \varsigma(A_1))  \wedge NC(tr, \varsigma(A_2))
  \rightarrow
\neg (\varsigma'(X^1_{A_3}) = \varsigma(X^1_{A_2}))
]
\]


\subsubsection{An authentication property}

Consider a two role protocol, such that role 1 finishes its execution after $n$ steps
and role 2 finishes its execution after $p$ steps.
For this kind of protocols we give a variant of the week
agreement property~\cite{lowe97hierarchy}.
Informally, this property states that whenever an instantiation of role
2 finishes, there exists an instantiation of role 1 that has finished
and they  agree on some value for some variable and they have indeed
talked to each other.
In our example we choose this variable to be $X^1_{A_1}$.
Note that we capture that some agent has finished its execution by
quantifying appropriately over the local states of that agent. 
More precisely, we quantify only over the states where it indeed has
finished its execution.
\begin{multline*}
\phi_a(tr) =
 \faLS{2,p}{tr}{\varsigma} \exLS{1,n}{tr}{\varsigma'}\\
[NC(tr, \varsigma(A_1))\wedge NC(tr, \varsigma'(A_2)) 
\rightarrow
 (\varsigma(X^1_{A_1}) = \varsigma'(X^1_{A_1})) \wedge 
(\varsigma(A_2) = \varsigma'(A_2)) \wedge (\varsigma(A_1) = \varsigma'(A_1))]
\end{multline*}

Notice that although in its current version our logic is not powerful enough to
specify stronger versions of agreement (like injective or bijective
agreement), it could be appropriately extended to deal with this more
complex forms of authentication.

\section{\uppercase{Main Result}}
\label{section-main}

Recall that our goal is to prove that  $\map{\Pi}\models
\map{\phi}\Rightarrow \Pi\models \phi$. However, as explained in the
introduction this property does not hold in general.
The following example sheds some light on the reasons that cause the
desired implication to fail.

\begin{example}\label{ex:counter}
Consider the first step of some protocol where  $A$ sends a
message to $B$ where some part is intended for some third agent.
\[
\begin{array}{rcl}
A& \rightarrow B:& \{N_a, \{N_a\}_{\ek(C)}, \{N_a\}_{\ek(C)}\}_{\ek(B)}
\end{array}
\]

The specification of the programs of $A$ and $B$ that corresponds to this
first step is as follows (in the definition below $C^1_{A_2}$ and
$C^2_{A_2}$ are variables of sort ciphertext).

\[
\begin{array}{lll}
\Pi(1) & = &  (\init, \encrypt{\ek(A_2)}{\langle X^1_{A_1},\langle 
\encrypt{\ek(A_3)}{X^1_{A_1}}^{\indexa{1}},
\encrypt{\ek(A_3)}{X^1_{A_1}}^{\indexa{2}}\rangle\rangle}^{\indexa{3}})\\[2ex]
\Pi(2) & = & 
(\encrypt{\ek(A_2)}{\langle X^1_{A_1},\langle C^1_{A_2}, C^2_{A_2}\rangle\rangle}^{L},
\stop)
\end{array}
\]
We assume that $A$ generates twice the message $\{N_a\}_{\ek(C)}$.
Notice that we stop the execution of $B$ after it receives the first
message since this is sufficient for our purpose, but its execution 
might be continued to form a more realistic example. 

Consider the security property $\phi_1$ that states that if $A$ and $B$
agree on the nonce $X^1_{A_1}$ then $B$ should have received twice the
same ciphertext.
\begin{multline*}
\phi_1(tr) = \faLS{1,2}{tr}{\varsigma} \faLS{2,2}{tr}{\varsigma'}\\
NC(tr, \varsigma(A_1)) \wedge NC(tr, \varsigma(A_2)) \wedge
(\varsigma(X^1_{A_1}) = \varsigma'(X^1_{A_1})) \rightarrow
(\varsigma'(C^1_{A_2}) = \varsigma'(C^2_{A_2}))
\end{multline*}
This property clearly does not hold for any normal execution of 
the labeled protocol since $A$ always sends ciphertexts with distinct
labels. Thus $\Pi\not\models \phi_1$.

On the other hand, one can show that
we have $\map{\Pi}\models \map{\phi_1}$ in the unlabeled execution model. 
Intuitively, this holds because if $A$ and $B$ are honest agents and
agree on $X^1_{A_1}$, then the message received by $B$ has been emitted by
$A$ and thus should contain identical ciphertexts (after having
removed their labels). 
\end{example}





\subsection{Logic $\Ltwo^l$}
\label{section-logic}

The counterexample above relies on the fact that two ciphertexts
that are equal in the model without labels may have been derived from
distinct ciphertexts in the model with labels.
Hence, it may be the case that although $\map{t_1}\neq
\map{t_2}\Rightarrow t_1\neq t_2$, the contrapositive implication 
$\map{t_1}=\map{t_2}\Rightarrow t_1= t_2$
does not
hold, which in turn entails that formulas that contain equality
tests between ciphertexts may be true in the model without labels, but
false in the model with labels. 
In this section we identify a fragment of $\Lone^l$, which we call
$\Ltwo^l$ where such tests are prohibited.
Formally, we avoid equality tests between arbitrary terms by
forbidding arbitrary negation over formulas and allowing equality tests
only between \emph{simple} terms.

\begin{definition}
A term $t$ is said \emph{simple} if $t\in \X.a\cup\X.n$ or $t=a$ for
some $a\in \id$ or $t= n(a, j, s)$ for some $a\in \id$, $j,s\in\N$.
\end{definition}
An important observation is that for any simple term $t$ it holds that
$\map{t}=t$.

\begin{definition}
The formulas of the logic $\Ltwo^l$ are defined as follows:
\[
\begin{array}{lll}
F(tr) & ::= & NC(tr, t_1) \db \neg NC(tr, t_1) \mid F(tr) \wedge F(tr)
\db F(tr) \vee F(tr) \mid (t_1 \neq t_2) \db (u_1 = u_2) \mid\\
 & & \faLS{i,p}{tr}{\varsigma}F(tr) \db \exLS{i,p}{tr}{\varsigma}F(tr),  \\
\end{array}
\]
where $tr\in \SymbTr$ is a parameter, $i,p \in \N$,
$t_1,t_2\in\tfs^l$ and $u_1,u_2$ are simple terms.
\end{definition}
Since simple terms also belong to $\tfs^l$, both equality and
inequality tests are allowed between simple terms.

The corresponding unlabeled logic $\Ltwo$ is defined as expected. Note
that $\Ltwo^l\subset \Lone^l$ and $\Ltwo\subset \Lone$.

\subsection{Theorem}
\label{section-theorem}

Informally, our main theorem says that to verify if a protocol satisfies
some security formula $\phi$ in  logic $\Ltwo^l$, it is sufficient to verify
that the unlabeled version of the protocol satisfies $\map{\phi}$. 
\begin{theorem}
Let $\Pi$ be a protocol and $\phi\in\Ltwo^l$, then
$\map{\Pi}\models\map{\phi} \Rightarrow \Pi\models\phi$.
\end{theorem}

\begin{proof}
Assume $\map{\Pi}\models\map{\phi}$.
We have to show that for any trace $tr\in\Exec(\Pi)$,
$\interp{\phi(tr)}=1$.
>From lemma \ref{lem4} it follows that 
$\map{tr} \in \Exec(\map{\Pi})$, thus 
$\interp{\map{\phi}(\map{tr})}=1$, since $\map{\Pi}\models\map{\phi}$.
Thus, it is sufficient to show that
$\interp{\map{\phi}(\map{tr})} \Rightarrow \interp{\phi(tr)}$.
The following lemma offers the desired property.
\end{proof}

\begin{lemma}
\label{lem3}
Let $\phi(tr)\in\Ltwo^l$ for some $tr\in\SymbTr$,
$\interp{\map{\phi}(\map{tr})}$ implies $\interp{\phi(tr)}$.
\end{lemma}

\begin{proof}
The proof of the lemma is by induction on the structure of
$\phi(tr)$.

\begin{itemize}
\item $\phi(tr) = NC(tr, t)$ or $\phi(tr) = \neg NC(tr, t)$.
$\interp{NC(tr, t)}=1$, if and only if $t\in \id$ and
  $t$ does not occur in a $\corrupt$ event for the trace $tr$.
 This is equivalent to $\map{t}\in \id$ and
  $\map{t}$ does not occur in a $\corrupt$ event for the trace $\map{tr}$.
Thus $\interp{NC(tr, t)}=1$ if and only if $\interp{\map{NC(tr, t)}} =
\interp{NC(\map{tr}, \map{t})}=1$.

\item $\phi(tr) = (t_1\neq t_2)$.
  We have that $\map{\phi}(\map{tr}) = (\map{t_1} \neq
  \map{t_2})$ holds.
   Assume by contradiction that $\phi(tr)$ does not hold, \textit{i.e}
  $t_1=t_2$. This implies $\map{t_1} = \map{t_2}$, contradiction.

\item $\phi(tr) = (u_1 = u_2)$ with $u_1,u_2$ simple terms.
  We have that $\map{\phi}(\map{tr}) = (\map{u_1} = \map{u_2})$
  holds. Since $u_1$ and $u_2$ are simple terms, we have $\map{u_i} = u_i$,
  thus $u_1 = u_2$. We conclude that $\phi(tr)$ holds.

\item The cases $\phi(tr) = \phi_1(tr) \vee \phi_2(tr)$ or $\phi(tr) = \phi_1(tr) \wedge
  \phi_2(tr)$ are straightforward.

\item $\phi(tr) = \faLS{i}{tr}{\varsigma} F(tr)$.
If $\map{\phi}(\map{tr})$ holds, this means that for all 
$(\theta, i, p)) \in \LS_{i, p}(\map{tr})$, 
$\interp{\map{F}(\map{tr})[\theta/\varsigma]} = 1$.

Let $(\theta', i, p) \in \LS_{i, p}(tr)$. We consider
$\interp{F(tr)[\theta'/\varsigma]}$. 
Since $tr \in \Exec(\Pi)$ implies 
$\map{tr} \in \Exec(\map{\Pi})$ (Lemma~\ref{lem4}), we have $(\map{\theta'}, i, p) \in \LS_{i, p}(\map{tr})$.
By induction hypothesis,
$\interp{\map{F}(\map{tr})[\map{\theta'}/\varsigma]} = 1$
implies that  $\interp{F(tr)[\theta'/\varsigma]} = 1$. It follows that
\[
\forall (\theta', i, p) \in \LS_{i, p}(tr)\; \interp{F(tr)[\theta'/\varsigma]} = 1.
\]
Thus, $\phi(tr)$ holds.

\item $\phi(tr) = \exLS{i}{tr}{\varsigma} F(tr)$.
If $\map{\phi}(\map{tr})$ holds, this means that there exists
$(\theta, i, p)) \in \LS_{i, p}(\map{tr})$, such that
$\interp{\map{F}(\map{tr})[\theta/\varsigma]} = 1$.

By definition of the mapping function, there exists $(\theta', i, p)
\in \LS_{i, p}(tr)$ such that $\map{\theta'}=\theta$.
By induction
hypothesis, $\interp{F(tr)[\theta'/\varsigma]} = 1$. Thus 
there exists $\theta'$, such that 
$\interp{F(tr)[\theta'/\varsigma]} = 1$. Thus, $\phi(tr)$ holds.
\end{itemize}
\end{proof}

\section{\uppercase{Discussion}}
\label{section-discussion}

We conclude with a brief discussion of two interesting aspects of our
result. 
First, as mentioned in the introduction, the only property needed for
our main theorem to hold is that the underlying deduction system
satisfies the condition in Lemma~\ref{lem1}, that is $S \deductl m
\Rightarrow \map{S} \deduct \map{m}$. 
In fact, an interesting result would be to prove a more abstract and
modular version of our theorem.  

Secondly, a natural question is whether  the converse of our main
theorem holds.   
We prove that this is not the case.
More precisely, we show that there exists a protocol $\Pi$ and a
property $\phi$ such that $\Pi\models \phi$ but $\map{\Pi}\not\models
\map{\phi}$. 
Let $\Pi$ be the protocol defined in Example~\ref{ex:counter}.
Consider a security property  $\phi_2$ that states on the
contrary that whenever $A$ and $B$ agree on the nonce $X^1_{A_1}$ then
$B$ should have received two distinct ciphertexts. 
Formally: 
\begin{multline*}
\phi_2(tr) = \faLS{1,2}{tr}{\varsigma} \faLS{2,2}{tr}{\varsigma'}\\
 NC(tr, \varsigma(A_1)) \wedge NC(tr, \varsigma(A_2)) \wedge
(\varsigma(X^1_{A_1}) = \varsigma'(X^1_{A_1})) 
\rightarrow
(\varsigma'(C^1_{A_2}) \neq \varsigma'(C^2_{A_2}))
\end{multline*}
where $C^1_{A_2}$ and $C^2_{A_2}$ are variables of sort ciphertext.

This property clearly does not hold for any honest execution of the
unlabeled protocol since $A$ always sends twice the same ciphertext,
and thus $\map{\Pi}\not\models\map{\phi_2}$.
On the other hand however,  one can show that this property holds for
labeled protocols since, if $A$ and $B$ are honest agents and agree on
$X^1_{A_1}$, it means that the message received by $B$ has been
emitted by $A$ and thus contains two distinct ciphertexts. Thus,
$\Pi\models \phi_2$. 
We conclude that, in general,  $\Pi\models \phi$ does not
imply $\map{\Pi}\models\map{\phi}$. 




\bibliographystyle{apalike}
\small
\bibliography{labels}

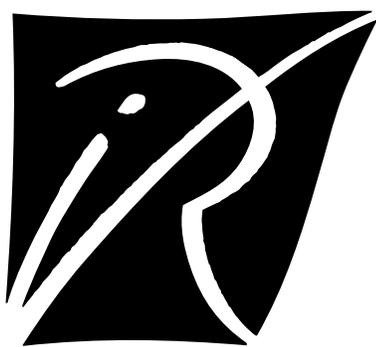
\end{document}